\documentclass{article}

\def\thebibliography#1{\section{References}\list
 {[\arabic{enumi}]}{\settowidth\labelwidth{[#1]}\leftmargin\labelwidth
 \advance\leftmargin\labelsep
 \usecounter{enumi}}
 \def\newblock{\hskip .11em plus .33em minus .07em}
 \sloppy\clubpenalty4000\widowpenalty4000
 \sfcode`\.=1000\relax}

\usepackage{enumitem}
\setlist{nosep, leftmargin=14pt}
\usepackage{spconf,amsmath,graphicx}
\usepackage{booktabs}
\usepackage{array}
\usepackage{tabularx}
\usepackage{mwe} 
\usepackage{multirow}
\usepackage{tabularx}
\newcolumntype{Y}{>{\centering\arraybackslash}X}
\usepackage{graphicx}
\usepackage{subcaption}
\usepackage{floatrow}
\usepackage{float}
\floatstyle{plaintop} 
\restylefloat{table}  
\usepackage{floatrow}
\usepackage{makecell}
\usepackage{xcolor}

\usepackage{adjustbox}

\title{Impact of regularization on achieved resolution in 3D tunable structured illumination microscopy (TSIM)}
 \name{Arash Atibi, Abdulaziz Alqahtani, Mohammed Younis, Chrysanthe Preza}
\address{
Department of Electrical and Computer Engineering, The University of Memphis, Memphis, TN, USA
}

\begin{document}
\maketitle
\begin{abstract}
We present a study that evaluates the impact of regularization on the achieved resolution in restorations from a {\it novel} three-dimensional (3D) Structured Illumination Microscopy (3D-SIM) system with desirable tunability properties. This contribution is the first performance evaluation of the Tunable SIM (TSIM) system through the restoration process. The study quantifies the achieved resolution in restorations, from simulated TSIM data of a 3D star-like object, at various expected resolution limits controlled by system parameters, and at different noise levels mitigated by the Generalized Wiener filter, a computationally efficient method, successfully applied to other conventional 3D-SIM systems. We show that theoretical  TSIM resolution limits are attained in the absence of noise, while with increasing noise levels, the necessary increase in regularization and residual restoration artifacts contributed to a $\sim$ 5\%-10\% and a 20\% reduction in the axial achieved resolution, in 20-dB and 15-dB data, respectively, which is within the pixel size (20 nm) limitation. 

\end{abstract}

\section{Introduction}
\label{sec:intro}
Three-dimensional (3D) Structured Illumination Microscopy (3D-SIM)  enables resolution enhancement in widefield microscopy by up to a factor of two \cite{Gustafsson}. Recent advancements in the field  \cite{Saavedra2023} introduced a  tunable SIM (TSIM) system, leveraging a unique combination of a quasi-monochromatic extended source and a Wollaston prism to achieve a tunable 3D structured illumination (SI) pattern. The TSIM system offers independent tunability of the lateral and axial resolutions, achieved by adjusting the lateral modulation frequency of the SI pattern and the source size, respectively. This is not possible in the three-wave SIM (3WSIM) \cite{Gustafsson} due to the non-linear coupling of the lateral and axial modulation frequencies of its SI pattern \cite{Saavedra2023}. Additionally, the TSIM system reduces data needs by 40\% compared to standard 3WSIM, requiring only 3 phase-shifted images per pattern orientation versus 5 in the 3WSIM scheme. Therefore, the TSIM system offers advantages over conventional SIM techniques. 
While the essential features of the TSIM system are elaborated upon in \cite{Saavedra2023}, an important aspect yet to be delved into is the impact of noise on the TSIM system's performance. Noise, inherent in all imaging systems, can significantly affect the accuracy 
of images restored by solving the inverse imaging problem. 

In this paper, we investigate the impact of noise on the performance of the TSIM system by employing the 3D Generalized Wiener Filter (3D-GWF) reconstruction process \cite{Gustafsson} using simulated data. This approach facilitates the evaluation of system performance by directly comparing the reconstructed images with the ground truth. Our choice of the 3D-GWF method is motivated by its early use in 3D-SIM image restoration and its computational efficiency, which allows observations to be made rapidly when simulation and restoration parameters are varied. Additionally, its widespread adoption by several commercial 3D-SIM instruments warrants a closer investigation of its efficacy in mitigating noise through regularization and the impact it has on the achieved super-resolution. This investigation enables us to assess the integration of the 3D-GWF method with the TSIM system. In a future study, we plan to explore the use of our 
model-based with a positivity constraint method, shown to be more accurate than the 3D-GWF, but also more computationally expensive \cite{Van:21}. 

In what follows, Section \ref{sec2} provides an overview of the 3D-SIM forward imaging model based on the 3D SI pattern generated in the TSIM system. In Section \ref{sec3}, we present numerical experiments to evaluate the impact of regularization on the achieved resolution. 
Finally, Section \ref{sec4} summarizes our findings and provides concluding remarks. 

\section{Background}
\label{sec2}

\subsection{Forward imaging model of 3D-SIM}
3D-SIM, as elucidated in \cite{8759445}, enables modeling the recorded 3D image intensity as a convolution of the fluorophore density distribution within the sample, $f(\boldsymbol{x}, z)$ modulated by the structured illumination (SI) pattern, and the point spread function (PSF) of the widefield microscope, $h(\boldsymbol{x}, z)$ as follows:
\begin{equation}\label{eq:2.1}
    g(\boldsymbol{x}, z)=\sum_{k=1}^{K} \left[f(\boldsymbol{x}, z) j_{k}(\boldsymbol{x})\right] \otimes \left[h(\boldsymbol{x}, z)i_k(z)\right],
\end{equation}
where $\boldsymbol{x}=(x, y)$ and $z$ are the transverse and the axial coordinates respectively;  $i_k(z)$ and $j_k(\boldsymbol{x})$ denote the separated axial and lateral functions of the SI pattern (oriented along the x-axis), respectively.

The illumination pattern of the TSIM system is described in \cite{Saavedra2023}, and it can be formulated as follows:
\begin{equation}\label{eq:2.2}
i(\boldsymbol{x},z)=1+\lvert V(z)\rvert\cos{\left(2\pi u_m\cdot \boldsymbol{x}+\phi+\Phi\left(z\right)\right)},
\end{equation}
where $V(z)$ is the visibility function, $u_m$ is the lateral modulation frequency, $\phi$ is the phase shift, and $\Phi(z)$ is the argument of the complex visibility $V(z)$. Eq. \eqref{eq:2.2} simplifies to $i(\boldsymbol{x}, z) = \sum_{k=1}^{3} j_k(x) i_k(z)$, where the individual lateral function $j_k(x)$ and axial function $i_k(z)$ are defined as follows:  both $i_1(z)$ and $j_1(x)$ are set to $1$; $i_2(z)$ and $i_3(z)$ take the form of $\lvert V(z)\rvert \cos(\Phi\left(z\right))$ and $-\lvert V(z)\rvert\,\sin(\Phi\left(z\right))$, respectively, while $j_2(x)$ and $j_3(x)$ correspond to $\cos(2\pi u_m \cdot \boldsymbol{x} + \phi)$ and $\sin(2\pi u_m \cdot \boldsymbol{x} + \phi)$, respectively.

\subsection{Restoration Method}\label{sec:3D-GWF-implementation}
In this paper, the restoration of 3D-SIM is carried out using the 3D-GWF, one of the first methods developed for 3D processing of 3D-SIM data \cite{Gustafsson}. The 3D-GWF follows a non-iterative four-step procedure, which includes decomposition, deconvolution, shifting, and recombination of the 3D-SIM components. This process requires several images to be obtained at 3 different orientations of the phase-shifted SI pattern so that the decomposition step can be achieved by solving a set of linear equations through a matrix inversion. For TSIM there are only 3 equations to be solved in each orientation of the pattern and therefore only 3 phase-shifted TSIM images are required \cite{Saavedra2023}. The deconvolution step includes a regularization parameter (the well-known Wiener parameter), chosen to effectively balance noise attenuation with the preservation of achieved resolution in the final image. A detailed exploration of the 3D-GWF method and strategies for choosing the regularization parameter in 3WSIM are provided in \cite{gustafsson2000surpassing} and \cite{KARRAS201969}, respectively.

\section{Numerical Experiments}\label{sec3}
\subsection{Simulation parameters}
The SI pattern of our 3D-TSIM system is characterized by an extended arbitrary source size and a distinct visibility function $V(z)$. When the illumination source is modeled as a rectangle function of dimension $L$,  its Fourier Transform is a \texttt{sinc} function, leading to the real-valued visibility function \cite{Saavedra2023}:
\begin{equation}
V(z) = \mathrm{sinc}\left(\frac{zu_mL}{nM_{\mathrm{i\ell\ell}}f_c}\right),
\end{equation}
where $\mathrm{sinc}(x) = \frac{\sin(\pi x)}{\pi x}$, $u_m$ is the lateral modulation frequency, $n = 1.515$ is the refractive index of the immersion medium of the objective lens, $M_{\mathrm{i\ell\ell}} = 0.0222$ is the illumination magnification, and $f_c = 100$ nm is the focal length of the collimation lens. 
The simulation parameters used in our study, include an emission wavelength $\lambda = 530$ nm and a numerical aperture, NA $= 1.4$, of the objective lens. For these parameters, the lateral and axial cutoff frequencies of the optical transfer function (OTF) of the native widefield microscopy (WFM) system are computed as $u_c = 2\,\text{NA}/\lambda=5.28\,\mu m^{-1}$ and $w_c = \frac{n-\sqrt{n^2-\text{NA}^2}}{\lambda}=1.77\,\mu m^{-1}$, respectively. 

In our study, we simulated TSIM data corresponding to three particular SI pattern modulation frequencies, namely, $u_m = 0.5u_c$, $0.75u_c$, and $0.8u_c$, matching the selection made in our previous study \cite{Saavedra2023}. For each modulation frequency, our analysis uses the largest source size, $L_{max}$ (Table~\ref{tab1}), that can be used for a given modulation frequency before the clipping effect, as discussed in \cite{Saavedra2023}, becomes a limiting factor. The lateral and axial resolution limits for WFM are computed using the expressions: $dx= 0.61 \lambda/\text{NA}=231$ nm and $dz=1/w_c=566$ nm, respectively. For this simulation study, the predicted TSIM lateral and axial resolution limits are computed using  $dx_{\text{SIM}}=dx/(1+u_m/u_c)$ and  $dz_{\text{SIM}}=dz/\left(\widetilde{w_c}/w_c\right)$, respectively. Here, $\widetilde{w_c}=w_c +\frac{L_{max}~u_m}{2nM_{\mathrm{i\ell\ell}}f_c}$ represents the effective axial cutoff frequency of the TSIM system \cite{Saavedra2023}. Table~\ref{tab1} displays the expected lateral and axial TSIM resolutions associated with each modulation frequency.
\begin{table}[ht!]
\centering
\renewcommand{\arraystretch}{0.5}
\caption{Simulation Parameters for Lateral ($dx_{\text{SIM}}$) and Axial ($dz_{\text{SIM}}$) Resolution Assessment.}
\begin{tabular}{ccccc}
\toprule
$u_m/u_c$ &  $L_{max}$ & $\widetilde{w_c}/w_c$ &$dx_{\text{SIM}}$ & $dz_{\text{SIM}}$ \\
\midrule
$0.50$ & $3.8$ mm & $1.845$ & $154$ nm & $307$ nm\\
$0.75$ & $2.7$ mm & $1.901$ & $132$ nm & $298$ nm \\
$0.80$ & $2.4$ mm & $1.854$ & $128$ nm & $305$ nm \\
\bottomrule
\end{tabular}
\label{tab1}
\end{table}
\begin{figure}[ht!]
    \centering
    \includegraphics[width=0.8\linewidth]{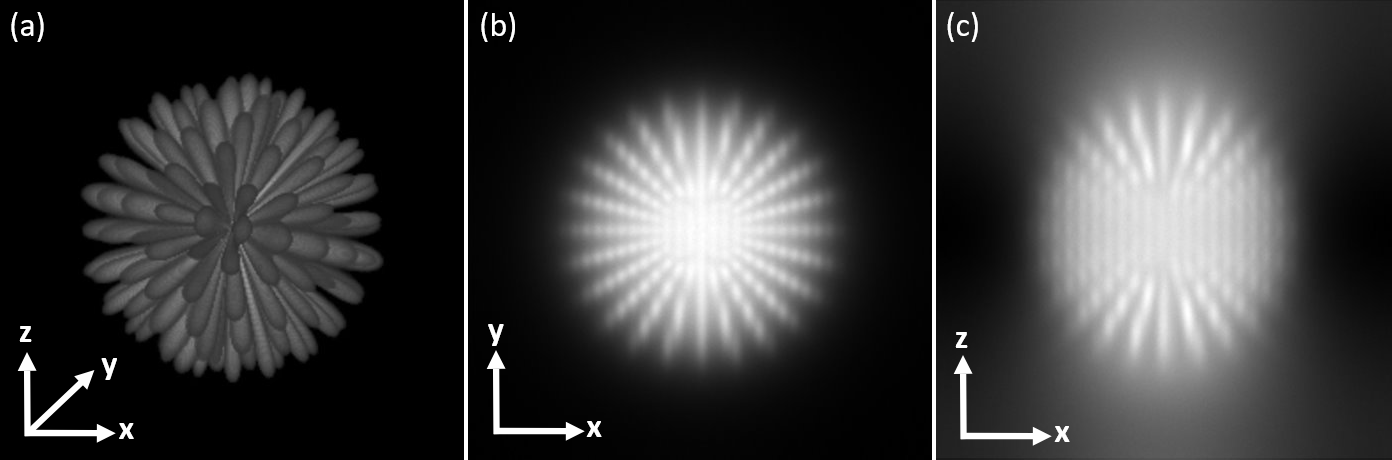}
    \caption{Numerical star-like object used in the simulations (a), and section images taken from the center of the raw simulated 3D TSIM data (SI pattern modulation frequency $u_m = 0.75u_c$), with a noise level of SNR = 20 dB: (b) XY-section image, and (c) XZ-section image.}
    \label{fig1}
\end{figure}
\begin{figure*}[ht!]
  \centering
  \includegraphics[width=14cm,height=10cm]{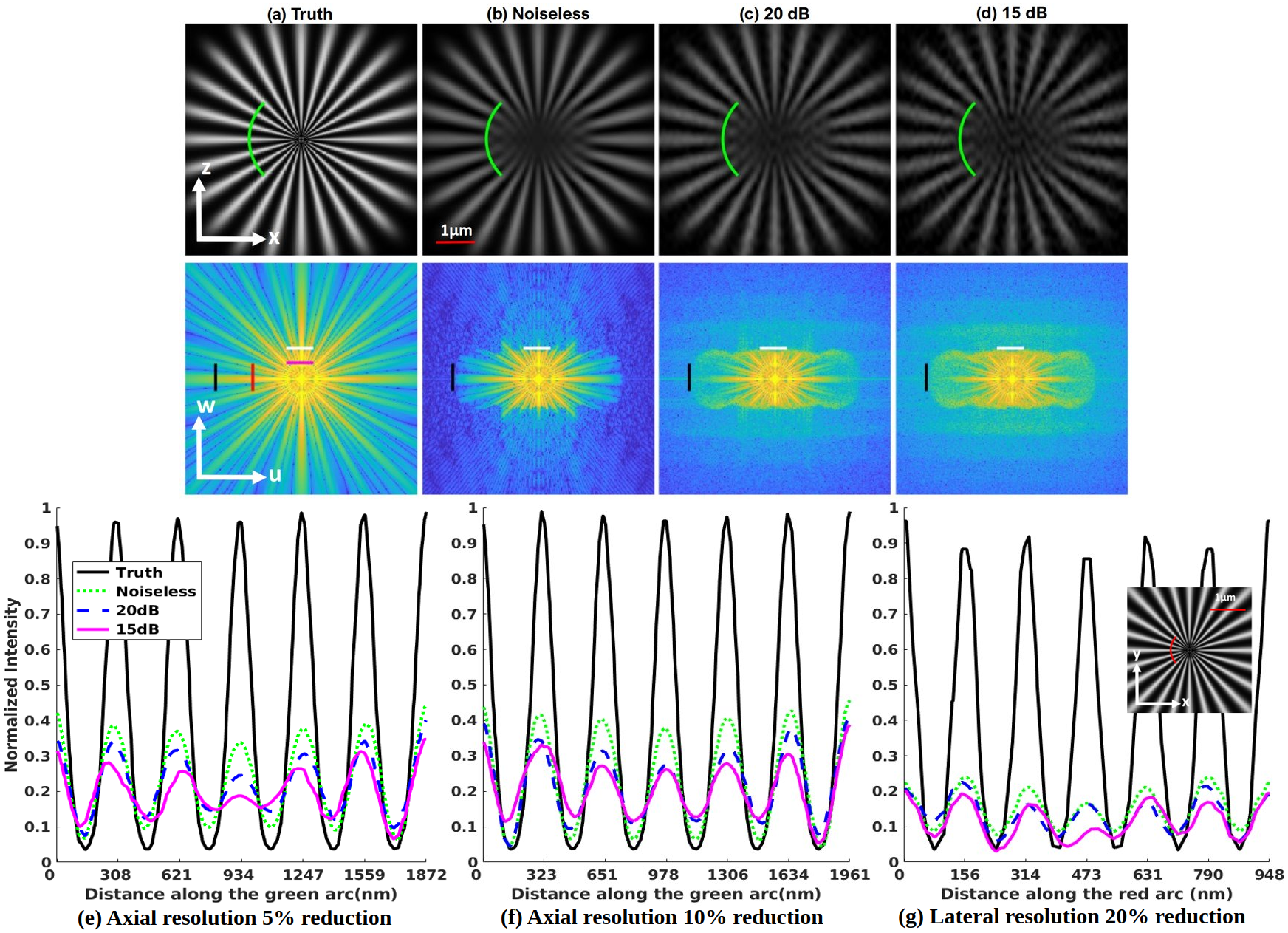}
 \caption{Axial and lateral resolution comparison and frequency analysis of true and restored images of the 3D star-like object (Fig.~1), achieved with the 3D-GWF method and the TSIM system using SI pattern modulation frequency $u_m = 0.75u_c$. XZ-section images from the center of the 3D images, with green arcs marking the theoretical axial resolution limit (first row), and uw-section images at the middle plane from the absolute value of the 3D Fourier transform (second row) of the true object (a), and 3D-GWF restorations from: (b) noiseless data with a regularization parameter (RP) of 1.00\,e$-$4; (c) noisy data with SNR = 20 dB  (RP = 5.50\,e$-$4); and (d) noisy data with SNR = 15 dB (RP =  1.00\,e$-$3). Normalized intensity profiles along a green arc corresponding to a reduction of the theoretical limit by: (e) 5\%, resulting in achieved axial resolution (AR) of $\sim 312$ nm; and  (f) 10\%, resulting in achieved AR of $\sim 327$ nm. (g) Normalized intensity profiles along the red arc shown in the XY-section image from the true object (inset image) and corresponding arcs in XY-section images from the restorations (images not shown). The achieved lateral resolution is $\sim 158$ nm (representing a 20\% reduction from the theoretical limit). The 3D-GWF restored volumes are $\ell_2$-normalized, with negative values set to zero, and images are displayed using the same color scale for a quantitative comparison (top row). Image intensities (second row) are displayed using a global maximum value and the same logarithmic scale of the form log(1+x). Red and magenta bars mark the distance from the origin at the theoretical lateral and axial cut-off frequencies of the WFM system in (a, second row). Black and white bars in (a-d, second row) mark the expected theoretical SIM cut-off frequencies based on the simulation conditions used.}
  \label{fig2}
\end{figure*}

To investigate the performance of the tunable 3D-SIM system using 3D restoration, we simulated the 3D TSIM image (Fig. 1b \& c) of an isotropic $3$D star-like object on a $512 \times 512 \times 512$ grid, with a $20$-nm voxel size (Fig. 1a). The simulated object is characterized by $6$ spokes in each quadrant ($24$ in total) with each spoke having a length of $\sim 3$ $\mu$m. We define the axial resolution limit as the center-to-center distance between two neighboring spokes along the green arc shown in the XZ-section image in Fig. 2(a).  Similarly, the lateral resolution limit is defined as the center-to-center distance between two neighboring spokes along the red arc shown in the inset XY-section image in Fig.~2(g). Using the above parameter settings and a 3D PSF computed using the Gibson and Lanni model \cite{Gibson:91}, we generated the simulated data using Eq. \eqref{eq:2.1}, which is then 
downsampled to a grid size of $256 \times 256 \times 256$ with a 40-nm voxel, to get the low-resolution data (Fig. 1b \& c). Eq. \eqref{eq:2.1} is computed 3 times using the SI pattern (Eqs. 2 \& 3) along three orientations in the XY-plane (at angles $ 0^{\circ}, 60^{\circ}, 120^{\circ}$) to achieve isotropic resolution, with 3 SI pattern phases $\phi$, shifted by a $(2\pi/3)$-rads step starting with $\phi = 0$ rads, at each orientation. In addition, Poisson noise was incorporated in the forward images at different levels resulting in noisy images with a signal-to-noise ratio (SNR) of 20 dB and 15 dB, computed from the mean of the squared root of the intensity at each pixel in the 3D image. 

\subsection{Summary of results} 

We investigated the performance of the 3D TSIM system through the restoration process using the 3D-GWF method, under various SI pattern modulation frequencies, noise levels, and regularization parameters. The regularization parameter was manually optimized for each case to obtain the best result. For noiseless data, we used a regularization parameter in the order of the one used in 20-dB data in each case, to demonstrate the impact of regularization alone on achieved resolution and overall performance. Several restorations using different values of the regularization parameter (within the range of $10^{-2}$ to $10^{-5}$) were evaluated using intensity profiles taken from the restoration to demonstrate the achieved axial (Fig.~2e \& f) and lateral (Fig.~2g) resolutions. In general, a 0.1 drop in intensity from peak to valley is required for the achieved axial resolution (see Fig.~2e \& 2f for the 15 dB case). The mean square error (MSE) and the structural similarity index measure (SSIM)~\cite{4775883}, computed between the true object and the restored images, were also used to compare globally the image restoration performance in each case. 

Selected results summarized in Table \ref{tab2}, show that the restoration performance metrics and the achieved resolutions, deteriorate with increasing noise levels, as expected. The best MSE and SSIM values were obtained for the $u_m = 0.5 u_c$ case, even though lower lateral and axial resolutions are achieved, as expected, than in the other two modulation frequencies (Table 1). The reason for this is a better restoration (higher contrast) of low frequencies that compensates for lower performance at higher ones. Results for the $0.75 u_c$ and $0.8 u_c$ cases show similar performance in terms of the MSE and SSIM. However, for the  $0.8 u_c$ case where the highest resolution is expected, the impact of higher noise and regularization is more significant, resulting in a higher reduction in the achieved resolutions compared to the other two cases. 


In Figure~\ref{fig2}, we further illustrate the results obtained for the $u_m = 0.75 u_c$ case by comparing XZ-section images from the numerical object (top row, Fig.~2a) and the 3D-GWF restorations from data at different noise levels (top row, Fig. 2(b),(c)\&(d)). Although the spokes of the star-like object in the restored images appear resolved at the green arcs, which denote the theoretical prediction for the axial resolution (= 298 nm), the intensity profiles taken along the arc, showed that the peak-to-valley drop was at least 0.1 (a metric we consider for achieved resolution) for most of the spokes, but not around the central one in restorations from noisy data. Intensity profiles taken at arcs corresponding to a 5\% decrease in the axial resolution showed that this problem persisted for the 15-dB result, while the resolution achieved is  $\sim 312$ nm in the 20-dB case (Fig.~2e); profiles taken at arcs corresponding to a 10\% decrease in the axial resolution demonstrate that the achieved resolution is $\sim 327$ nm in the 15-dB case (Fig.~2f).  Similarly, Fig.~2(g) shows that the achieved lateral resolution is $\sim 158$ nm, which represents approximately a 20\% decrease in the lateral resolution. The frequency analysis of the restorations obtained at the different noise levels, confirms that the highest frequencies present in the restorations are very close to the predicted theoretical limits for the cutoff frequencies marked by the white and black bars (Fig. 2, second row), but they are corrupted by the noise amplification through the 3D-GWF process in the noisy restorations (Fig. 2c \& d, second row).

\begin{table}[htbp]
\centering
\caption{Image Reconstruction Quality Metrics and Achieved Resolution for Various Modulation Frequencies and Noise Levels using the following Regularization Parameters: $10^{-4}$ for Noiseless, $5.5\cdot10^{-4}$ for $20$ dB and $10^{-3}$ for $15$ dB.}
\label{tab2}
\begin{adjustbox}{width=\columnwidth,center}
\begin{tabular}{cccccc}
\toprule
\multirow{2}{*}{$\displaystyle\frac{u_m}{u_c}$} & SNR & MSE & SSIM & \multicolumn{2}{c}{\makecell{Achieved Resolution (nm)\\ \& Reduction Percentage}}  \\
& (dB) & &  & Lateral & Axial  \\
\midrule
\multirow{3}{*}{0.5} & $\infty$ & $7.86\cdot10^{-4}$ & 89.3 & 169, 10\% & 307, 0\% \\
 & 20 & $1.04\cdot10^{-3}$ & 85.2 & 176 , 15\% & 321, 5\% \\
& 15 & $1.31\cdot10^{-3}$ & 80.6 & 184, 20\% & 337, 10\% \\
\midrule
\multirow{3}{*}{0.75} & $\infty$ & $9.90\cdot10^{-4}$ & 85.8 & 144, 10\% & 298, 0\% \\
 & 20 & $1.24\cdot10^{-3}$ & 81.6 & 151, 15\% & 312, 5\% \\
 & 15 & $1.57\cdot10^{-3}$ & 75.9 & 158, 20\% & 327, 10\% \\
\midrule
\multirow{3}{*}{0.8} & $\infty$ & $9.95\cdot10^{-4}$ & 85.6 & 140, 10\% & 305, 0\% \\
 & 20 & $1.27\cdot10^{-3}$ & 81.1 & 147, 15\% & 335, 10\% \\
 & 15 & $1.61\cdot10^{-3}$ & 75.2 & 164, 25\% & 365, 20\% \\
\bottomrule
\end{tabular}
\end{adjustbox}
\end{table}

\section{CONCLUSION}\label{sec4}
The study presented here focuses on the robustness of the 3D Tunable Structured Illumination Microscopy (TSIM) system in the presence of noise, considering restorations obtained through the 3D-GWF method and using simulated data. The trade-off between resolution enhancement and noise mitigation was explored at different noise levels and various theoretically expected TSIM resolution values, controlled by the tunable lateral modulation frequency of the system and the size of the illuminating source.  
The restoration results presented here (and their global assessment through the MSE and SSIM metrics) are highly influenced by the OTF values inside its support, and not only by its cutoff frequencies, based on which theoretical resolution limits are computed. 

Our restoration results confirm the capability of the TSIM system to provide tunable resolution enhancements. In the absence of noise, the theoretically expected lateral (within a 10\% decrease)  and axial resolution improvements were attained with minimal regularization, which is used to mitigate the inversion of small values of the OTF, in the deconvolution step of the 3D-GWF method. In the presence of increasing noise, the use of increased regularization and residual restoration artifacts contributed to a reduction of the achieved resolutions by 5-25\% from theoretical limits (Table 2), which is within the pixel size (20 nm) limitation. Our metric for resolution is the spacing between 2 neighboring spokes in the object, which for $u_m = 0.8 u_c$ is only  6 and 15 pixels for the lateral and axial resolutions, respectively, making lateral resolution harder to assess due to poor sampling. 

The study contributes to understanding the efficacy of the 3D-GWF integrated with the TSIM system, paving the way for further exploration and optimization using different model-based methods to solve the TSIM inverse imaging problem.

\section{Acknowledgments}
\label{sec:acknowledgments}
The authors wish to thank Dr. Genaro Saavedra (Optics Department, University of Valencia, Spain) for valuable discussions that contributed to this project.


\begin{thebibliography}{99}

\bibitem{Gustafsson}
M.~G.~L.~Gustafsson, L.~Shao, P.~M.~Carlton, C.~J.~R.~Wang, I.~N.~Golubovskaya, W.~Z.~Cande, D.~A.~Agard, and J.~W.~Sedat, ``Three-dimensional resolution doubling in wide-field fluorescence microscopy by structured illumination," \textit{Biophys. J.}, vol. 94, no. 12, pp. 4957--4970, 2008.



\bibitem{Saavedra2023}
A.~Gimeno-Gomez, S.~P.~Dajkhosh, C.~T.~S.~Van, J.~C.~Barreiro, C.~Preza, and G.~Saavedra, ``Management of the axial modulation of the illumination pattern in structured illumination microscopy using an extended illumination source," \emph{Opt. Express}, vol. 31, no. 22, pp. 36568--36589, Oct. 2023. 

\bibitem{Van:21}
C.~T.~S.~Van and C.~Preza, ``Improved resolution in 3D structured illumination microscopy using 3D model-based restoration with positivity-constraint," \emph{Biomed. Opt. Express}, vol. 12, no. 12, pp. 7717--7731, Dec. 2021. 

\bibitem{8759445}
H. Shabani, S. Labouesse, A. Sentenac, and C. Preza, ``Three-dimensional deconvolution based on axial-scanning model for structured illumination microscopy," in \textit{Proc. IEEE 16th International Symposium on Biomedical Imaging (ISBI 2019)}, pp. 552--555, 2019.
\bibitem{gustafsson2000surpassing}
M. G. L. Gustafsson, ``Surpassing the lateral resolution limit by a factor of two using structured illumination microscopy," \emph{J. Microsc.}, vol. 198, no. 2, pp. 82--87, 2000. 

\bibitem{KARRAS201969}
C. Karras, M. Smedh, R. Förster, H. Deschout, J. Fernandez-Rodriguez, and R. Heintzmann, ``Successful optimization of reconstruction parameters in structured illumination microscopy – A practical guide," \emph{Opt. Commun.}, vol. 436, pp. 69-75, 2019. 

\bibitem{Gibson:91}
S. F. Gibson and F. Lanni, ``Experimental test of an analytical model of aberration in an oil-immersion objective lens used in three-dimensional light microscopy," \emph{J. Opt. Soc. Am. A}, vol. 8, no. 10, pp. 1601--1613, Oct. 1991. 

\bibitem{4775883}
Z. Wang and A. C. Bovik, ``Mean squared error: Love it or leave it? A new look at Signal Fidelity Measures," \emph{IEEE Signal Process. Mag.}, vol. 26, no. 1, pp. 98-117, 2009. 

\end{thebibliography}

\end{document}